Performance of Multi-group DIF Methods in Assessing Cross-Country Score Comparability
of International Large-Scale Assessments

**Dandan Chen**
Department of Educational Psychology
The University of Illinois at Urbana-Champaign
1310 S. Sixth St., Champaign, IL 61820-6925

## Introduction

Standardized large-scale testing can be a debatable topic, in which test fairness sits at its very core. With high stakes in ensuring equity and access to educational and professional opportunities, test fairness often concerns *bias* that disproportionally treats different population groups, distinguished by gender, race, or ethnicity (Lee & Zhang, 2017). It is "the one of greatest importance to the public" (Hambleton et al., 1991, p. 109). Modern methods to investigate item bias are mostly based on the framework of *differential item functioning* (DIF; Berk, 1982; Holland & Wainer, 1993), according to Ayala (2009) and Penfield and Lam (2000). *DIF* refers to the situation in which "different groups of test takers with similar overall ability, or similar status on an appropriate criterion, have, on average, systematically different responses to a particular item" (p. 16). It is a formal definition of DIF provided by the American Educational Research Association, American Psychological Association, and National Council on Measurement in Education (2014). It is preferable to the term *bias*, as it is free from value-based judgment (Angoff, 1993). *Uniform DIF* exists when test takers from one group consistently have a better chance of answering correctly than test takers of the same competency level from another group. *Nonuniform DIF* occurs when this relationship is not consistent (Mellenbergh, 1982; as cited in Swaminathan & Rogers, 1990).

The vast majority of the literature on DIF methods is limited to two groups (instead of more than two). It means a critical gap in the measurement science of important practical significance to assessment practice as different methods might result in different test items being flagged with DIF. Specifically, in international large-scale assessments (ILSAs), which involve a multitude of countries, different multi-group DIF methods might cast doubt onto the cross-country comparability of ILSA scores in opposite directions. Note that currently the score rankings from ILSAs such as Program for International Student Assessment (PISA; OECD, 2019a) are defining the global competitiveness of nations and affecting education policymaking (Fischman et al., 2019). Also, DIF analysis, as part of the routine work on ILSAs, requires high precision and consistency to ensure the ILSA scores are truly comparable across countries. It is critical therein that the assessment researchers and practitioners can agree on a multi-group DIF method with high power to correctly capture DIF when analyzing the ILSA score data from many countries.

Two out of five recent multi-group DIF detection methods were found capable of capturing both the uniform and nonuniform DIF that affects test fairness. Still, no prior research has demonstrated the relative performance of these two methods when they are compared with each other. These two methods are the improved Wald test and the generalized logistic regression procedure. Calling for attention to cross-country comparability of ILSA scores, this paper specifically concentrated on the performance of two multi-group DIF methods in capturing DIF with ILSA score data from more than two countries. It sought to answer the following question: How different are the results from the two multi-group DIF detection methods that can capture both the uniform and nonuniform DIF? Note that the "groups" in this project were defined as examinee groups differentiated by country.

This study started with reviewing the existing literature to identify what multi-group DIF methods have been developed and studied to detect DIF with data from more than two groups. In this process, five recent multi-group DIF detection methods were found having been developed, and two of them can capture both the uniform and nonuniform DIF. Also, a research gap was found in the relative performance of two recent multi-group DIF methods when they are compared with each other. Then, the two multi-group DIF methods were assessed by checking the commonalities and differences between two sets of empirical results from the methods in analyzing the same TIMSS math score data. These two multi-group DIF methods were the improved Lord's $\chi^2$ test or Wald test with the Wald-1 or Wald-2 linking algorithm (Cai et al., 2011; Langer, 2008; Kim et al., 1995; as cited in Woods et al., 2012), and the generalized logistic regression procedure (Magis et al., 2011). The former was referred to as "improved Wald test" in short throughout this paper.

The results unveiled that the generalized logistic regression procedure detected a lot more DIF items than the improved Wald test and that its detected DIF items kept increasing as the number of groups increased. The latter might be partially explained by the feature of inflated Type I error rates when more groups are taken into consideration in the generalized logistic regression, as some researchers have specified (Magis et al., 2011). One key takeaway from this study is that the improved Wald test is relatively more established than the generalized logistic regression procedure for multi-group DIF analysis, despite its relatively limited sensitivity to uniform DIF. These findings were expected to inform the selection of a multi-group DIF method for ILSA test score analysis, demonstrating that the choice of the multi-group DIF method can impact the results.

## Literature Review

### Measurement Invariance and DIF in ILSA

Today, ILSAs are mostly based on Item Response Theory (IRT; Richardson, 1936; Lawley, 1943; Tucker, 1946), in which *invariance* is an essential assumption, partly undergirded by the assessment of *DIF*. The conceptual definition of invariance is that "some properties of a measure should be independent of the characteristics of the person being measured, apart from those characteristics that are the intended focus of the measure" (Millsap, 2007, p. 462). Assessing DIF is indispensable to establishing the point that the invariance assumption is met in IRT-based ILSAs.

An operational definition of invariance is that "the probability of an observed score, given the attribute level and the group membership, is equal to the probability of that given only the attribute level" (Mellenbergh, 1989; Meredith, 1993; as cited in Croudace & Brown, 2012). Note that, when applied to test items, this operational definition of invariance is precisely the definition of DIF (Croudace & Brown, 2012), as described in Introduction. It implies the equivalence of the two and the necessity of assessing DIF across countries for ensuring the legitimacy of comparing IRT-based scores across countries as ILSAs often do.

ILSAs are highly susceptible to DIF across countries, however. The source of invariance, or DIF, in testing settings, engage variables not that relevant to test takers. They include item characteristics (e.g., item format, item content) and contextual variables (e.g., classroom size, teaching practices); both are attributable to differences between testing situations (Zumbo, 2007). Both exist in ILSAs, which are often translated or adapted to diverse contexts (Grisay et al., 2009), such as the socio-cultural situations across countries. This nature implicates ILSAs' high susceptibility to DIF, and it is critical to ensure the validity of the common cross-country comparisons using ILSA scores by assessing the actual presence of DIF among ILSA items.

### DIF Detection Methods

The most commonly used methods to assess DIF assume the data in analysis are unidimensional, and they can be classified into two general approaches (Magis et al., 2011; Langer, 2008). One approach, called the *IRT approach*, superimposes an IRT model on the provided data. The Lord's $\chi^2$ or Wald test (Lord, 1980; Wald, 1943), the likelihood-ratio test (Thissen et al., 1988), and the latent class logistic regression (Zumbo et al., 2015) follow this approach. The other approach, called the *observed-score approach*, does not require an IRT model. The Mantel-Haenszel procedure (MH; Mantel & Haenszel, 1959), logistic regression (Swaminathan & Rogers, 1990), and SIBTEST (Shealy & Stout, 1993) fall into the second category. For the IRT approach, there are two subcategories of methods; one focuses on identifying differences in estimated item parameters between groups, whereas the other on identifying differences in the estimated areas between item response function curves from different groups (Kim et al., 1995). A thorough review of these methods can be found in Osterlind and Everson (2009) and Lee (2015).

A variety of DIF methods are available for complex data structures. In contexts with multidimensional data sets, multidimensional DIF methods are available. They include the multidimensional SIBTEST (MULTISIB; Stout et al., 1997) and the multidimensional DFIT (Oshima et al., 1997). Sophisticated DIF methods proposed more recently include the ones based on hierarchical models (e.g., Swanson et al., 2002), mixture models (e.g., Frederickx et al., 2010) and generalized linear mixed models (e.g. De Boeck, 2008). Studies like Bechger and Maris (2015) have gone further, proposing to examine DIF by focusing on DIF of item pairs instead of individual items, for parameter estimates of an item are not identified from observations, but relative estimates from item pairs are.

It is worth noting that these methods are either originally designed to compare two groups (e.g., Swaminathan & Rogers, 1990; Bechger & Maris, 2015), or have not yet established the evidence that they apply to more than two groups (e.g., De Boeck, 2008). Five recent DIF methods have been demonstrated with acceptable performance for analyzing DIF across multiple groups in the unidimensional framework. They are the generalized Mantel-Haenszel method (Penfield, 2001), generalized Lord's $\chi^2$ test or Wald test (Kim et al., 1995) with the Wald-1 linking algorithm (Cai et al., 2011; Langer, 2008; as cited in Woods et al., 2012), generalized logistic regression procedure (Magis et al., 2011), multiple-indicator multiple-cause modeling (MIMIC; Muthén, 1985, 1989), and multiple-group confirmatory factor analysis (MGCFA; Asparouhov & Muthén, 2014). Note that the generalized Maten-Haenszel method can only detect the uniform DIF, whereas the other four methods can detect both the uniform and nonuniform DIF. Also, the MIMIC method was found not straightforward enough for implementation when considering nonuniform DIF (Woods et al., 2012). As for MGCFA, the criteria typically recommended for evaluating DIF may not always be appropriate when a large number of groups are involved for comparison (Rutkowski & Svetina, 2014), which is the case with ILSAs.

A few existing studies have compared some of the aforementioned multi-group DIF methods in terms of their performance for capturing DIF, with simulated data or local assessment data, not the real-world data from major ILSAs. For example, Finch (2015) completed a Monte Carlo comparison of the generalized Mantel-Haenszel test, the generalized logistic regression, MGCFA, and the traditional Lord's chi-square test (Kim et al., 1995), using simulated data for two, three and six groups respectively. What looked more relevant to this study was Magis et al.'s comparison of the generalized logistic regression procedure and the generalized Mantel-Haenszel method. The results from the two methods were compared in capturing DIF across years of assessment scores from college students in Quebec Province in Canada.

Meanwhile, it is worth noting that some other multi-group DIF detection methods are not seen as ideal. For example, the IRT-LR (Thissen et al., 1986) was found producing a few less accurate standard deviations and slightly less power (given a small sample size) than the improved Wald test with Wald-1 algorithm. Also, the improved Wald test with Wald-2 algorithm has inflated Type I error rates as it does not designate anchor items, recommended only when it is used to select anchor items for the Wald-1 test (Woods et al., 2012). The LR lasso method has inflated Type I error rates and weak sensitivity, with no gain in statistical power (Rollins, 2018).

**DIF Detection in ILSAs**

Technical reports of the psychometric equivalence in assessments across countries, cultures or languages should include some systematic information about the extent to which such equivalence was achieved, though perfect psychometric equivalence can be hardly achieved (e.g., Grisay et al., 2009; Hambleton et al., 2004). While the assessment of DIF has become a routine part of test design (e.g., OECD, 2019b), the use of DIF detection procedures in major ILSAs is not well-documented. These ILSAs include Program for International Student Assessment (PISA; OECD, 2019a), Program for International Assessment of Adult Competencies (PIAAC; OECD, 2019b), Trends in International Mathematics and Science Study (TIMSS; IEA, 2017a), and Progress in International Reading Literacy Study (PIRLS; IEA, 2018).

For example, the technical report on PIAAC (OECD, 2019b) mentions DIF testing, but their explanation of the DIF analyses is not clear regarding what methods were employed for assessing DIF. TIMSS (IEA, 2016), PIRLS (IEA, 2017b) and PISA (OECD, 2015) have used IRT-based models to estimate item parameters and detect the difference in the parameter estimates across countries or test modes. However, the primary problem is with the insufficient description of specific algorithms employed in their DIF analyses. Not to mention, there is a lack of information concerning the relative performance of the method they employed as compared to other DIF methods existing in the literature, which is important for supporting their choice of methods. These facts pointed to a problem that is the efficiency of the DIF methods adopted in ILSAs remains a question, and it is not open to public investigation due to the limited documentation.

Some research articles have investigated DIF across more than two groups. A recent systematic review of peer-reviewed articles in 2000–2015 on PISA (Hopfenbeck et al., 2018) identified many sources that have investigated DIF across language groups. The DIF methods employed in these sources included the aforementioned MGCFA and logistic regression method. In addition, they covered a couple of novel DIF methods: the Linear Logistic Test Model (LLTM; Fischer, 1973), which assessed the group-by-facet *differential facet functioning* (DFF; Engelhard, 1992); the mixture Rasch model (MRM; Rost, 1990), which examined the *latent DIF* that exists between latent examinee groups; and the Multidimensional Random Coefficient Multinomial Logit Model (MRCMLM; Meulders & Xie, 2004), which is a hybrid model. As for DIF across countries, von Davier, Xu and Carstensen (2011) employed a general latent variable model to specify and compare two types of multidimensional IRT models for longitudinal data and found significant country-level DIF in ILSAs. Also, using MGCFA, Wu, Li and Zumbo (2007) demonstrated strong within-culture invariance and weak between-culture invariance in 21 cross-country comparisons with TIMSS math scores. More recently, Oliveri and von Davier (2017) adopted the discrete mixture distribution model (von Davier & Yamamoto, 2004), which is a latent mixture version of the mixed Rasch model, and found the invariance assumption was not fully met when examining the item parameter estimates across 30 countries and across test administrations using PISA scores.

The vast majority of the existing research that has investigated DIF in ILSA scores across countries is limited by the DIF methods applicable to two groups only. Cheema (2019) detected widespread uniform and nonuniform DIF in PISA data on science that show gender gaps in science literacy among both OECD and non-OECD countries. This multi-group analysis employed the logistic regression approach (Swaminathan & Rogers, 1990), which is limited to two-group comparisons. This applied study is a faulty practice and demonstrates a lack of understanding of DIF methods. Feskens, Fox, and Zwitser (2019) adopted the differential item pair analysis (Bechger & Maris, 2015), which is also limited to two-group DIF detection. It evaluated the extent to which PISA math items are subject to DIF, targeting the DIF between two groups that correspond to the two modes of current PISA tests.

In sum, this literature review led to two important findings. First, at least five multi-group DIF methods have been developed to assess DIF between observed examinee groups at the item level in the unidimensional framework. This study would want to concentrate on these methods instead of more complicated DIF methods for the sake of parsimony. The five methods are the generalized Mantel-Haenszel method (Penfield, 2001), generalized Lord's $\chi^2$ test or Wald test (Kim et al., 1995) with the Wald-1 linking algorithm (Cai et al., 2011; Langer, 2008; as cited in Woods et al., 2012), generalized logistic regression procedure (Magis et al., 2011), multiple-indicator multiple-cause modeling (MIMIC; Muthén, 1985, 1989), and multiple-group confirmatory factor analysis (MGCFA; Asparouhov & Muthén, 2014). Some of them were built on commonly recognized two-group DIF methods, with improved algorithms to minimize errors and increase reliability in DIF testing.

Second, two of these multi-group DIF methods outperform others, for they can capture both the uniform and nonuniform DIF, but no study has compared them with either simulated data or real-world assessment data. These two methods are the improved Wald test (Woods et al., 2012) and the generalized logistic regression procedure (Magis et al., 2011). Note that, as mentioned above, the generalized Mantel-Haenszel method can only detect the uniform DIF, the MIMIC method was not straightforward enough with the nonuniform DIF detection, and the MGCFA may not have the appropriate criteria for the DIF detection with a large number of groups, which is the case with ILSAs. Future research that fills in this gap can shed light on the selection of a multi-group DIF method for the ILSA score analysis.

## Methodology

At this point, findings from the literature review informed the design of this study. They concern which multi-group DIF method is the best for accurately capturing both the uniform and nonuniform DIF. The ensuing empirical analyses assessed the relative performance of the improved Wald test and the generalized logistic regression procedure by comparing results from the two. Similarities or differences in the detected DIF items were expected to demonstrate how the performances of the two methods converge or diverge, and to inform the method selection for future multi-group DIF analysis in ILSAs.

### Data

The data for this analysis were from the math section of TIMSS 2015, which originally estimated Grade-8 students' math and science proficiency scores for 56 countries and 6 participating entities for benchmarking. The number of test takers ranged from 2,074 to 18,012 (IEA, 2017b) for individual countries or participating entities. Each student was randomly administered one of the 14 TIMSS assessment booklets that covered a series of math and science items, in the form of multiple-choice items or constructed-response items (IEA, 2017a).

These items corresponded to four content-related domains, namely Algebra, Data and Chance, Geometry, and Number. Meanwhile, they corresponded to three cognitive-concentrated domains, namely Applying, Knowing, and Reasoning (IEA, 2017a). Item-level descriptive information and parameter estimates across countries in addition to the item correct rate and almanacs aggregated by country can be found in the international database of TIMSS 2015's official webpage (TIMSS & PIRLS International Study Center, 2019a).

Test items of TIMSS 2015 were scaled with three IRT models, respectively. They are the two-parameter model (2PL; Lazarsfeld, 1950) for TIMSS's constructed-response items having been scored as correct or incorrect; three-parameter model (3PL; Hambleton & Swaminathan, 1985) for the multiple-choice items having been scored as correct or incorrect; and a generalized partial credit model (GPCM; Muraki, 1992) for constructed-response items scored with more than two options (IEA, 2017b).

Booklet 13 was randomly selected from the 14 booklets, covering 29 items in math. Based on the item type information provided by TIMSS & PIRLS International Study Center (2019b), 2PL and 3PL unidimensional IRT models are appropriate for all but one of these items. One test item required the GPCM; for the convenience of modeling, this item was excluded from analysis. Hence, the data set contained a total of 28 test items using only 2PL and 3PL models for further analysis. There were eight items in Algebra, five items in Data and Chance, six items in Geometry, and nine items in Number.

Considering the need for sufficient sample size for DIF analysis, the countries selected for this study had a sample size among the highest and a score data structure that reasonably met the assumption of unidimensionality. A total of six countries were selected to enable the check of differences in results from the two DIF methods when involving varied numbers of country groups, namely three, four, and six country groups.

**Dimensionality Testing**

Both of the multi-group DIF methods targeted in this study, namely the improved Wald test and the generalized logistic regression procedure, assumed unidimensionality. Hence, to start with, the assumption of unidimensionality was assessed for the selected data. Specifically, factor analysis was performed to understand the dimensionality of the data by country using Mplus. One-factor, two-factor, and three-factor models were fitted in the dimensionality assessment. Eigenvalues, AIC/BIC, RMSEA, CFI, and SRMR indices were computed and compared across models.

**DIF Analyses**

**Improved Wald test.** The improved Wald test is an upgraded version of Kim et al.'s (1995) generalized Wald test for multi-group analysis, adapted from Lord's (1980) Wald test. The "improvement" was achieved by employing the supplemented expectation maximization (SEM) algorithm (Cai, 2008; Meng & Rubin, 1991), instead of the joint maximum likelihood method. Following Woods et al.'s recommendations (2012), this improved Wald test starts with identifying anchor items with the Wald-2 algorithm, which flexMIRT refers to as the "sweep" procedure. Then, it proceeds with assessing the DIF using the Wald-1 algorithm for all except the anchor item identified from the first step, which flexMIRT names the "candidate" procedure.

Empirical and simulation studies (Langer, 2008; Woods et al., 2012) have demonstrated that the Wald-2 test can help control Type I error rates, estimating the covariances and parameters with higher accuracy than the original Wald test. Requiring no anchor item, the Wald-2 test first fits a model in which the mean and standard deviation of the reference group are fixed at 0 and 1, respectively. Doing so helps identify the scale and estimate the means and standard deviations of all focal groups. In this process, all the item parameters are constrained to be equal across groups. Then, it fits a model that fixes the means and standard deviations of focal groups using the values obtained for focal groups in the first step, while allowing all item parameters to vary freely between groups. A covariance matrix of item parameter estimates is used for calculating Kim et al.'s (1995) chi-square statistic, shown as

$$Q_i = (\boldsymbol{C}\boldsymbol{v_i})^T (\boldsymbol{C}\textstyle\sum_i \boldsymbol{C})^{-1}(\boldsymbol{C}\boldsymbol{v_i}) \tag{1}$$

where

> $\boldsymbol{C}$ is a matrix of contrast coefficients that identify how parameters are compared across groups;
> $\boldsymbol{v_i}$ is a matrix that holds the differences between item parameter estimates for all groups;
> $\sum_i$ is a matrix that holds covariances between item parameter estimates for all groups.

The hypothesis testing relies on this chi-square statistic $Q_i$. The degree of freedom for testing is the number of parameters compared for each item across groups. The related hypothesis concerns the homogeneity of item parameters (Kim et al., 1995), written as

$$H_0\colon \boldsymbol{C}\boldsymbol{\varepsilon_i} = \boldsymbol{0} \tag{2}$$

where

> $\boldsymbol{C}$ is a matrix of contrast coefficients that identify how parameters are compared across groups;
> $\boldsymbol{\varepsilon_i}$ is a vector of item parameters.

As for the Wald-1 test, different from the Wald-2 test, it requires anchor items and fits only one model. Using this model, the Wald-1 test fixes the mean and standard deviation of the reference group at 0 and 1 respectively. It then estimates the mean and standard deviation of focal groups in addition to item parameters simultaneously. In this process, item parameters are either constrained to be equal across groups (for anchor items) or freed to vary between groups (for non-anchor items). A simulation study (Woods et al., 2012) has demonstrated that the Wald-1 test has superior performance when compared with the Wald-2 test and the IRT-LR test (Thissen et al., 1988). Woods et al. did not recommend the Wald-2 test unless it is used for selecting anchor items. Thus far, no study has compared the improved Wald test with the generalized logistic regression procedure.

Since both the Wald-1 and Wald-2 tests assume an IRT-based model and the final data set of this study contained 2PL and 3PL models, 2PL and 3PL models were stipulated respectively for relevant test items in flexMIRT. Moreover, for the 3PL items, a prior was specified for their c parameters in both the Wald-1 and Wald-2 tests. This prior was extracted from a log-normal distribution with a mean of -1.1 and a standard deviation of 0.5. The expected a posteriori (EAP; Bock & Aitkin, 1981) method was employed to estimate the IRT scale scores.

Following the Mean p-value Selection approach (MP; Kopf et al., 2015), the item that displays the highest mean p-value from the auxiliary DIF test (i.e., the Wald-2 test herein) was selected as the anchor item. Note that MP was chosen as the anchor method because it is one of the three new anchor selection methods that have been demonstrated to outperform some classic anchoring strategies, such as the all-other anchor method (Woods, 2009) and the Number of Significant Threshold (NST) selection strategy (Wang, 2004), according to Kopf et al. (2015). The Type I error rate for hypothesis testing of the significance of DIF was fixed at the .05 level.

**Generalized logistic regression procedure.** Distinct from the improved Wald test, the generalized logistic regression method does not require an IRT model to fit with the score data. ~~Also, it can be used when there is no reference group.~~ It is an extension of Swaminathan and Rogers's (1990) logistic regression method ~~for two-group comparisons~~. Its DIF model is expressed as (Magis et al., 2011)

$$\log \frac{\pi_{ig}}{1-\pi_{ig}} = \alpha + \beta S_i + \alpha_g + \beta_g S_i \tag{3}$$

where

> $\pi_{ig}$ is the probability that Respondent *i* from Group g (g=0, 1, 2, ..., F) answers an item correctly;

$\alpha$ is the common intercept parameter shared by all groups;

$\beta$ is the common slope parameter shared by all groups;

$\alpha_g$ is the intercept parameter of Group $g$;

$\beta_g$ is the slope parameter of Group $g$;

$S_i$ is the test score of Respondent $i$.

This formula can be rewritten as

$$\log\frac{\pi_{ig}}{1-\pi_{ig}} = \begin{cases} \alpha + \beta S_i & if \ g = 0 \\ (\alpha + \alpha_g) + (\beta + \beta_g)S_i & if \ g \neq 0 \end{cases} \tag{4}$$

It shows the intercept and slope parameters are equal to $\alpha$ and $\beta$ respectively in the reference group ($g = 0$), and they are equal to $(\alpha + \alpha_g)$ and $(\beta + \beta_g)$ respectively in the focal groups.

This parameterization can be utilized in the hypothesis testing for testing nonuniform DIF (NUDIF) and uniform DIF (UDIF), specified respectively as

$$H_0: \beta_1 = \beta_2 \ldots = \beta_F = 0 \qquad\qquad (NUDIF) \tag{5}$$
$$H_0: \alpha_1 = \alpha_2 = \cdots = \alpha_F = 0 \ | \ \beta_1 = \beta_2 = \cdots = \beta_F = 0 \quad (UDIF) \tag{6}$$

Equation (5) specifies that nonuniform DIF is present when at least one slope parameter $\beta_g$ is different from 0. Equation (6) stipulates that uniform DIF is present when at least one intercept parameter $\alpha_g$ is different from 0 on condition that there is no nonuniform DIF (i.e., all slope parameters equal 0). A vector of parameters from Equation (4) is then estimated using the maximum likelihood estimation method (MLE; Agresti, 1990).

It is worth noting its likelihood ratio test. Two nested models are compared for the generalized logistic regression procedure in this test. One for the null hypothesis and the other for the alternative hypothesis. For testing the presence of NUDIF, the null and alternative hypotheses are shown as (Magis et al., 2011)

$$\begin{aligned} M_0 &= \log\frac{\pi_{ig}}{1-\pi_{ig}} = (\alpha + \alpha_g) + \beta S_i \\ M_1 &= \log\frac{\pi_{ig}}{1-\pi_{ig}} = (\alpha + \alpha_g) + (\beta + \beta_g)S_i \end{aligned} \qquad \begin{array}{c} \textit{for NUDIF} \\ (7) \end{array}$$

For testing the presence of UDIF, the two hypotheses are specified as

$$\begin{aligned} M_0 &= \log\frac{\pi_{ig}}{1-\pi_{ig}} = \alpha + \beta S_i \\ M_1 &= \log\frac{\pi_{ig}}{1-\pi_{ig}} = (\alpha + \alpha_g) + \beta S_i \end{aligned} \qquad \begin{array}{c} \textit{for UDIF} \\ (8) \end{array}$$

A maximized likelihood can be computed for the two models respectively for either NUDIF and UDIF, marked as $L_0$ and $L_1$. Then, Wilks' (1938) lambda statistic $\Lambda$, which supports the decision-making in the hypothesis testing for NUDIF and UDIF, can be computed based on these likelihoods. This statistic is also known as the likelihood ratio statistic, expressed as

$$\Lambda = -2\log(\frac{L_0}{L_1}) \tag{9}$$

In this test, a large Wilks' lambda statistic $\Lambda$ implicates the presence of NUDIF or UDIF.

To date, this generalized logistic regression procedure has not been studied with actual ILSA assessment data that involve numerous countries. Nor has it been compared with the improved Wald test. Its performance was examined with simulated data, which found the number of groups did not affect the performance of this method, and that Type I error rates were inflated in non-DIF conditions (Svetina & Rutkowski, 2014). This method has been applied to analyze a regional assessment data set for assessing score comparability across test administrations (Magis et al., 2011).

In this study, the generalized logistic regression procedure was run via the "difR" package in RStudio. An item purification procedure (Clauser & Mazor, 1998; Candell & Dragsow 1988) was enabled via R to identify DIF items while running the DIF analysis, with no need to stipulate an anchor item beforehand. In this process, after an initial DIF analysis, items detected with DIF were removed, and the DIF analysis was re-run. Then this step is repeated until the immediate ensuing iteration returns the same DIF classifications.

It is worth noting issues concerning the statistical significance of the parameter estimates from this method. The "difR" package in RStudio provides only two options for evaluating this significance, using either the likelihood ratio approach inherent to the IRT-LR test (Thissen et al., 1988) or the traditional Wald test (Wald, 1939, 1943). The likelihood ratio test was used for this estimation, for it has been found more reliable than the traditional Wald test (e.g., Langer, 2008; Agresti, 2002). Also, while the significance level here was fixed at .05 as in the improved Wald test, the estimated p-values from DIF analysis were revised via the Holm and Benjamini-Hochberg adjustments (Holm, 1979), one of the recommended adjustment methods concerning DIF (Kim & Oshima, 2013). This study considers such adjustments as necessary whenever it is technically possible for they help control Type I errors in multiple comparisons. The "difR" package provides this adjustment option, and not to mention, the prior research has noted the inflated Type I error rates from the generalized logistic regression procedure as with the increase of the number of groups in DIF analysis (Magis et al., 2011).

## Results

### Descriptive Statistics

The final TIMSS data selected for analysis came from six countries, namely Canada (CAN; n=620), Australia (AUS; n=729), United States (USA; n=735), New Zealand (NZL; n=566), Thailand (THA; n=464), and Iran (IRN; n=441). Table 1 specifies, for each item, its IRT-based model predefined by TIMSS, its missing rate, point-biserial correlation coefficient, and proportion correct. In total, there are 17 items scaled with a 3PL model (i.e., Items 1-3, 6-11, 13-16, 20, 23, 25-26), and 11 items with a 2PL model (i.e., Items 4-5, 12, 17-19, 21-22, 24, 27-28). The point-biserial correlations of the score data from the six countries were beyond .35 for all except Item 3 (r=.31) and Item 8 (r=.19), suggesting most of the items in this data set had reliable score data. When analyzing this data by country, the point-biserial correlation coefficients varied from .01 to .73, of which the coefficients were below .30 for Items 1, 3, 8, 14, 16, 17, and 25. After omitting missing values, the average correct rate ranged approximately from .19 to .77 for individual items, and it ranged from .08 to .84 when looking at the data segregated by country. These statistics showed large disparity across items in the correct rate, implicating a mixture of easy and difficult test items.

### Data Structure

Results from the dimensionality testing supported the use of a unidimensional model for the selected data. The ratio of first to second eigenvalues ranged from 4.9 to 8.5 across the six countries, indicating a strong primary dimension. While AIC values favored a larger number of dimensions, BIC values consistently favored a unidimensional model. Using common rules of thumb (MacCallum et al., 1996; Bentler & Bonett, 1980; Hu & Bentler, 1999), RMSEA values were below .05 for the one-factor model for all countries, CFI values were above .95 for the one-factor model for all countries except Canada that has a slightly lower value when fitted with a two-factor model, and SRMR values were below .08 for all countries. Table 2 specifies these model fit statistics by country.

**Inferences from the DIF Analyses**

Tables 3–6 report DIF statistics from the two multi-group DIF methods. Canada was randomly selected out of the six countries as the reference group, with other countries being the focal groups in the three-, four- and six-group DIF analyses, of which the results were reported below. In the meantime, seven-, eight- and nine-group analyses were performed, which did not run with the generalized logistic regression procedure, and possible explanations were presented in the Discussion section.

Tables 5–6 report the group-specific parameter estimates and standard errors of the DIF items from the Wald-1 test and the generalized logistic regression method, respectively. Figure 1 plots the group-specific item characteristic curves specific for items detected with DIF via both the two multi-group DIF methods. These plots show the country-specific item characteristic curves were positioned relatively differently when using the correct response rate and the score information from the two multi-group DIF methods, respectively.

**Three-group DIF analysis.** The three countries selected for this analysis were Canada, the United States, and Australia. From the Wald-2 test with the data, Item 20 was identified as the anchor item, for it had the highest average p-value from DIF analysis. As shown in Table 3, the ensuing two multi-group DIF analyses respectively flagged different items with DIF. The Wald-1 test detected DIF in four items (i.e., Items 2, 4–5, and 19), and five items with nonuniform DIF (i.e., Items 2, 5, 16, 19, and 26). The generalized logistic regression procedure identified 16 items with DIF (i.e., Items 1–3, 5–8, 15–19, 24–26, and 28), nine items with nonuniform DIF (i.e., Items 2–3, 5, 7–8, 15–16, 21, and 23), and two items with uniform DIF (i.e., Items 5, 8). Apparently, the results from the two multi-group DIF methods were quite different.

**Four-group DIF analysis.** The four countries used in this analysis were Canada, the United States, Australia, and New Zealand. With the anchor item being Item 20 again for the same reason mentioned above, as reported in Table 3, the follow-up Wald-1 test detected DIF in three items (i.e., Items 2, 16, and 19), of which two items were also detected with DIF in the prior three-group Wald-1 test. Meanwhile, this test identified five items with nonuniform DIF (i.e., Items 2, 16–17, 19, and 26), of which all except one were detected with DIF in the prior three-group Wald-1 test. The generalized logistic regression procedure found 15 items with DIF (i.e., Items 1–3, 5–8, 16–19, 24–26, and 28), one less than the 16 items detected with DIF from the three-group analysis. Furthermore, it identified 15 items with nonuniform DIF (i.e., Items 1–3, 5–8, 16–19, 24–26 and 28), of which six items were also detected from the three-group analysis; and two items with uniform DIF, same as the two items identified from the three-group analysis (i.e., Items 5, 8). It was evident that most of the items identified with DIF in the six-group analysis using the generalized logistic regression procedure were found with DIF in the three-group analysis.

**Six-group DIF analysis.** The six countries covered in this analysis were Canada, the United States, Australia, New Zealand, Thailand, and Iran. Table 4 shows that, with the anchor item being Item 20 again, the Wald-1 test identified two items with DIF (i.e., Items 2 and 17), and six items with nonuniform DIF (i.e., Items 2, 16–19, and 26), of which five items were flagged in the prior four-group analysis. The generalized logistic regression procedure found 22 items with DIF, which left only six items without DIF, and all the items detected with DIF in the prior four-group analysis were flagged here again with DIF. Moreover, this analysis marked all except five items with nonuniform DIF, covering all with nonuniform DIF from the three-group analysis. Meanwhile, it identified nine items with uniform DIF, including all the items identified with uniform DIF in the three-group analysis. The massive increase in the number of items detected with DIF using the generalized logistic regression procedure was astounding when comparing the results from this analysis with prior analyses that involved fewer groups.

Overall, it is worth noting several critical features of the two sets of results from the three-, four- and six-group analyses using the two multi-group DIF methods. First, there was an overlap in the items detected with DIF between the two sets of results. All the items flagged with DIF from the Wald-1 test were also detected with DIF from the generalized logistic regression procedure. Also, the same applied to all but two items flagged with significant nonuniform DIF. Second, the items marked with DIF kept changing as the number of groups changed for each of the two multi-group DIF methods. For example, with the Wald-1 test, Items 4–5 marked with DIF in the three-group DIF analysis were not detected with DIF or nonuniform DIF in the four- or six-group analysis. With the generalized logistic regression, Item 15 detected with DIF in the three-group analysis was not flagged with DIF in the six-group analysis. Third, the two methods appeared to have different sensitivity in detecting uniform DIF. No item was detected with uniform DIF in all the DIF analyses with the Wald-1 test, whereas many items were detected with uniform DIF with the generalized logistic regression procedure.

One evident disparity between the two sets of results was in the total number of items detected with DIF, regardless of the number of involved groups. First, the number of DIF items detected with DIF using the generalized logistic regression procedure was much larger than that using the Wald-1 test, despite the overlap. The total number of items with DIF is 4 versus 16 from the Wald-1 test and the generalized logistic regression procedure respectively for the three-group analysis; this contrast is 3 versus 15 items in the four-group analysis, and 2 versus 22 items in the six-group analysis. Second, as the number of groups increased for the analysis, the total number of DIF items from the two multi-group DIF methods tended to change in opposite ways. As the number of countries increased from three to four and to six, the total number of items detected with DIF from the Wald-1 test decreased from four to three to two, suggesting the increased sample size or degrees of freedom brought by more countries in the analysis might affect the detection of DIF. Contrarily, when using the generalized logistic regression procedure, this total number increased from 16 to 22 as the number of countries groups for analysis increased from three to six, despite the drop of one DIF item when this number increased from three to four.

## Discussion

Which set of the DIF results from the two multi-group DIF methods truly comply with the actual differences between item-level responses? One might consider a simulation study for future research to provide an answer. For this study, given the empirical results and different scholars' perspectives as described above altogether, the improved Wald test was found relatively more established than the generalized logistic regression procedure, despite its relatively limited sensitivity in detecting uniform DIF. Researchers and practitioners in assessment might want to consider applying the improved Wald test in assessment work concerning binary item-level score data from more than two classes, schools, grades, countries, or test administrations.

How might the seemingly inflated DIF presence from the generalized logistic regression procedure be explained? One key feature of the two sets of empirical results was that a lot more items were detected with DIF using the generalized logistic regression procedure, compared with that with the Wald-1 test. Also, it is worth noting that, with the generalized logistic regression procedure, the total number of DIF items kept increasing with the number of country groups in analysis. Moreover, the seven-, eight-, and nine-group analyses with the generalized logistic regression procedure did not converge, and that the non-converging estimates showed all the items were detected with DIF. As Magis et al. (2011) have noted, with the generalized logistic regression procedure, more items would be flagged with DIF due to inflated Type I error rates when more groups are considered. It contradicts with Svetina and Rutkowski's (2014) finding from a simulation that the number of groups did not affect the performance of this method. The findings about the inflated DIF presence from this study appeared to support Magis et al.'s comment instead of Svetina and Rutkowski', although the Holm and Benjamini-Hochberg adjustment method was used to control for Type I error rates in the DIF analyses in this generalized logistic regression procedure.

Some evident limitations were inherent to the research design of this study. First, despite the recognition of the problem of the likelihood ratio test, this study used the likelihood ratio test instead of the improved Wald test for estimating parameters in the generalized logistic regression procedure. Second, a frequentist approach was employed instead of a Bayesian approach when estimating item parameters in the generalized logistic regression procedure. In effect, the availability of relevant software programs limited the method choices. The author was not aware of any software programs that enable the generalized logistic regression procedure with the improved Wald algorithms or the Bayesian approach in parameter estimation, and can only make the most use of the software options that were available.

Future research in the multi-group DIF analysis could evolve in several ways. One might start by confirming the patterns detected in the two sets of empirical results from the two multi-group DIF methods would remain true with other data sets of different structures. Demonstration with simulated data would be useful in this regard. Also, one might want to unpack the practical significance of item-level DIF to the actual ratings from ILSAs. Researchers have found the item-level DIF matter might be balanced out in the domain-level scoring in PISA's problem-solving tests (e.g., Oliveri & Ercikan, 2011; Oliveri et al., 2012). Similar research on TIMSS scores would be useful. If this case holds for TIMSS and other ILSAs, item-level DIF might practically pose a minor threat to the validity of ILSA scores that often materialize in the form of domain-level ratings. Moreover, to improve the generalized logistic regression procedure, one might want to examine and resolve the inflated Type I error issue as described above. At least two future directions have been made clear by this study. First, integrate the Wald-1 test into the generalized logistic regression programming. Second, employ a Bayesian paradigm (e.g., Frederickx et al., 2010) to calculate posterior DIF probabilities and effect sizes that could counterbalance inflated error rates, as Magis et al. (2011) have suggested.

More importantly, future research that examines method bias or construct bias beyond item bias as reflected by DIF of ILSAs could hold great promise. As exhibited by the current literature, item bias and its associated DIF techniques have received much more attention than method bias and construct bias, which is unfortunate as the latter two are also integral to the infringement on test fairness. Essential methods for assessing method bias include the monotrait-multimethod procedure, factor analysis, and analysis of covariance structures (van de Vijver & Poortinga, 1997). Methods for examining construct bias include the "cultural decentering" to investigate the correspondence of the instrument to definitions of a concept (Werner & Campbell, 1970). Future research that work on techniques in these two aspects could support a more comprehensive investigation of the test fairness of ILSAs.

In the end, it is vital to keep in mind the meaning of DIF and take the results from either of the two selected multi-group DIF methods with a grain of salt. Specifically, in terms of the statistical difference in ILSA scores between countries, DIF concerns the between-country divergence in the probability of answering an item correctly. As mentioned earlier, items with DIF are not necessarily biased against certain groups, for this DIF presence might only correspond to true difference. Similarly, items having not been detected with DIF are not necessarily unbiased, for the bias that might exist in this case do not materialize in true difference. Whether the captured DIF reflects true difference or bias in test design is a judgment call that needs additional information beyond the item-response data for DIF analyses. As well put by Hambleton and Swaminathan (1985), one may or may not be able to alleviate this true difference in assessment practice, and either case is beyond the purpose of a test or a measure itself.

Tables

**Table 1**

*TIMSS 2015: Statistics of Items by Country*

| Item ID | | Model | % of Missing Values | All Selected Data | | CAN | | USA | | AUS | | NZL | | THA | | IRN | |
|---|---|---|---|---|---|---|---|---|---|---|---|---|---|---|---|---|---|
| | | | | r | p | r | p | r | p | r | p | r | p | r | p | r | p |
| 1 | M052063 | 3PL | 0.2% | 0.37 | 0.39 | 0.28 | 0.35 | 0.44 | 0.48 | 0.37 | 0.31 | 0.44 | 0.31 | 0.55 | 0.40 | 0.42 | 0.50 |
| 2 | M052072 | 3PL | 0.2% | 0.47 | 0.56 | 0.36 | 0.51 | 0.49 | 0.61 | 0.51 | 0.6 | 0.46 | 0.57 | 0.61 | 0.46 | 0.51 | 0.56 |
| 3 | M052092 | 3PL | 0.3% | 0.31 | 0.26 | 0.23 | 0.18 | 0.50 | 0.46 | 0.31 | 0.23 | 0.42 | 0.20 | 0.08 | 0.15 | 0.08 | 0.23 |
| 4 | M052146A | 2PL | 0.2% | 0.43 | 0.54 | 0.47 | 0.64 | 0.39 | 0.49 | 0.44 | 0.53 | 0.50 | 0.51 | 0.43 | 0.56 | 0.46 | 0.50 |
| 5 | M052146B | 2PL | 0.2% | 0.54 | 0.20 | 0.48 | 0.31 | 0.50 | 0.15 | 0.55 | 0.16 | 0.24 | 0.68 | 0.14 | 0.57 | 0.21 |
| 6 | M062067 | 3PL | 1.0% | 0.44 | 0.68 | 0.43 | 0.79 | 0.37 | 0.78 | 0.43 | 0.68 | 0.42 | 0.67 | 0.49 | 0.52 | 0.36 | 0.58 |
| 7 | M062242 | 3PL | 1.6% | 0.51 | 0.60 | 0.48 | 0.65 | 0.47 | 0.76 | 0.55 | 0.62 | 0.58 | 0.61 | 0.46 | 0.43 | 0.28 | 0.37 |
| 8 | M062341 | 3PL | 1.2% | 0.19 | 0.29 | 0.09 | 0.21 | 0.47 | 0.56 | 0.13 | 0.2 | 0.18 | 0.20 | 0.03 | 0.24 | 0.01 | 0.26 |
| 9 | M052161 | 3PL | 0.3% | 0.47 | 0.77 | 0.40 | 0.84 | 0.43 | 0.80 | 0.49 | 0.81 | 0.45 | 0.79 | 0.44 | 0.65 | 0.42 | 0.62 |
| 10 | M052418A | 3PL | 0.3% | 0.58 | 0.41 | 0.57 | 0.54 | 0.57 | 0.39 | 0.59 | 0.42 | 0.62 | 0.42 | 0.59 | 0.36 | 0.47 | 0.29 |
| 11 | M052418B | 3PL | 0.4% | 0.43 | 0.55 | 0.40 | 0.61 | 0.39 | 0.59 | 0.4 | 0.55 | 0.47 | 0.57 | 0.48 | 0.49 | 0.40 | 0.43 |
| 12 | M062072 | 2PL | 3.1% | 0.55 | 0.57 | 0.39 | 0.71 | 0.47 | 0.64 | 0.5 | 0.63 | 0.55 | 0.66 | 0.66 | 0.32 | 0.53 | 0.29 |
| 13 | M062120 | 3PL | 5.6% | 0.52 | 0.54 | 0.45 | 0.62 | 0.54 | 0.56 | 0.52 | 0.62 | 0.52 | 0.54 | 0.49 | 0.39 | 0.46 | 0.39 |
| 14 | M052046 | 3PL | 0.3% | 0.35 | 0.36 | 0.29 | 0.44 | 0.32 | 0.37 | 0.41 | 0.4 | 0.33 | 0.38 | 0.36 | 0.28 | 0.21 | 0.24 |
| 15 | M052082 | 3PL | 0.3% | 0.45 | 0.54 | 0.42 | 0.67 | 0.45 | 0.52 | 0.38 | 0.58 | 0.40 | 0.56 | 0.47 | 0.38 | 0.57 | 0.46 |
| 16 | M052083 | 3PL | 0.3% | 0.44 | 0.39 | 0.46 | 0.57 | 0.39 | 0.41 | 0.44 | 0.44 | 0.51 | 0.33 | 0.38 | 0.29 | 0.09 | 0.18 |
| 17 | M062192 | 2PL | 2.7% | 0.37 | 0.19 | 0.28 | 0.16 | 0.40 | 0.30 | 0.34 | 0.12 | 0.27 | 0.08 | 0.62 | 0.25 | 0.58 | 0.21 |
| 18 | M062250A | 2PL | 1.5% | 0.52 | 0.53 | 0.52 | 0.63 | 0.56 | 0.47 | 0.43 | 0.65 | 0.44 | 0.62 | 0.57 | 0.49 | 0.54 | 0.23 |
| 19 | M062250B | 2PL | 1.9% | 0.57 | 0.31 | 0.52 | 0.49 | 0.56 | 0.29 | 0.53 | 0.36 | 0.57 | 0.29 | 0.64 | 0.17 | 0.51 | 0.16 |
| 20 | M052024 | 3PL | 0.3% | 0.55 | 0.55 | 0.51 | 0.66 | 0.56 | 0.57 | 0.57 | 0.6 | 0.53 | 0.57 | 0.52 | 0.40 | 0.44 | 0.40 |
| 21 | M052058A | 2PL | 0.2% | 0.54 | 0.70 | 0.47 | 0.82 | 0.47 | 0.76 | 0.47 | 0.75 | 0.46 | 0.74 | 0.60 | 0.54 | 0.61 | 0.47 |
| 22 | M052058B | 2PL | 0.2% | 0.59 | 0.28 | 0.53 | 0.40 | 0.55 | 0.32 | 0.59 | 0.31 | 0.56 | 0.28 | 0.73 | 0.15 | 0.55 | 0.09 |
| 23 | M052125 | 3PL | 0.3% | 0.55 | 0.44 | 0.43 | 0.54 | 0.52 | 0.50 | 0.57 | 0.49 | 0.55 | 0.46 | 0.43 | 0.26 | 0.67 | 0.28 |
| 24 | M052229 | 2PL | 0.2% | 0.46 | 0.46 | 0.49 | 0.45 | 0.43 | 0.52 | 0.51 | 0.42 | 0.54 | 0.32 | 0.60 | 0.41 | 0.46 | 0.63 |
| 25 | M062001 | 3PL | 0.4% | 0.37 | 0.53 | 0.38 | 0.68 | 0.35 | 0.61 | 0.31 | 0.54 | 0.43 | 0.46 | 0.31 | 0.47 | 0.21 | 0.35 |
| 26 | M062146 | 3PL | 0.6% | 0.44 | 0.40 | 0.35 | 0.35 | 0.43 | 0.38 | 0.55 | 0.49 | 0.58 | 0.40 | 0.42 | 0.38 | 0.43 | 0.38 |
| 27 | M062154 | 2PL | 0.8% | 0.61 | 0.57 | 0.57 | 0.70 | 0.57 | 0.57 | 0.64 | 0.63 | 0.63 | 0.62 | 0.55 | 0.47 | 0.57 | 0.38 |
| 28 | M062214 | 2PL | 0.5% | 0.56 | 0.44 | 0.47 | 0.55 | 0.49 | 0.40 | 0.57 | 0.53 | 0.56 | 0.49 | 0.66 | 0.27 | 0.54 | 0.34 |
| | % of Missing Values | | | 0.9% | | 1.7% | | 1.0% | | 0.9% | | 0.4% | | 0.5% | | 0.8% | |

*Note.* The column "p" displays the average correct rate, and the column "r" displays the point-biserial correlations (or item-total correlations).

**Table 2**

*Model Fit Statistics from Dimensionality Analysis*

| Country | Model | Eigenvalue | AIC | BIC | RMSEA | CFI | SRMR | $\Delta\chi^2$ | p-value |
|---------|-------|-----------|-----|-----|-------|-----|------|-----------|---------|
| CAN | 1-factor | 8.67 | 20231.33 | 20567.99 | 0.05 | 0.93 | 0.08 | | |
| | 2-factor | 1.76 | 20171.58 | 20627.85 | 0.04 | 0.95 | 0.07 | 156.18 | 0.00 |
| | 3-factor | 1.55 | 20157.35 | 20728.78 | 0.04 | 0.97 | 0.06 | 118.36 | 0.00 |
| USA | 1-factor | 10.36 | 23583.65 | 23887.24 | 0.03 | 0.97 | 0.06 | | |
| | 2-factor | 1.35 | 23564.72 | 23992.51 | 0.03 | 0.98 | 0.05 | 109.80 | 0.00 |
| | 3-factor | 1.24 | 23548.88 | 24096.26 | 0.03 | 0.98 | 0.05 | 88.55 | 0.00 |
| AUS | 1-factor | 10.96 | 22690.72 | 23007.54 | 0.04 | 0.96 | 0.07 | | |
| | 2-factor | 1.56 | 22643.96 | 23084.76 | 0.03 | 0.98 | 0.06 | 181.54 | 0.00 |
| | 3-factor | 1.46 | 22609.78 | 23169.96 | 0.02 | 0.99 | 0.05 | 163.29 | 0.00 |
| NZL | 1-factor | 11.53 | 17000.49 | 17295.51 | 0.04 | 0.97 | 0.07 | | |
| | 2-factor | 1.58 | 16956.52 | 17368.68 | 0.03 | 0.98 | 0.06 | 128.94 | 0.00 |
| | 3-factor | 1.34 | 16926.33 | 17451.30 | 0.03 | 0.99 | 0.05 | 82.38 | 0.00 |
| THA | 1-factor | 12.32 | 13705.77 | 13987.28 | 0.04 | 0.98 | 0.07 | | |
| | 2-factor | 1.44 | 13695.34 | 14088.63 | 0.03 | 0.99 | 0.06 | 87.43 | 0.00 |
| | 3-factor | 1.29 | 13689.56 | 14190.48 | 0.03 | 0.99 | 0.06 | 65.08 | 0.00 |
| IRN | 1-factor | 10.20 | 13409.49 | 13675.28 | 0.04 | 0.96 | 0.08 | | |
| | 2-factor | 1.53 | 13372.00 | 13748.20 | 0.03 | 0.98 | 0.07 | 88.99 | 0.00 |
| | 3-factor | 1.49 | 13349.97 | 13832.48 | 0.02 | 0.98 | 0.06 | 68.78 | 0.00 |

*Note.* $\Delta$ means the change in statistics across models.

**Table 3**

*TIMSS 2015: DIF Statistics from Two Methods (Three & Four Groups)*

| Item | Model | 3-group analysis | | | | | | 4-group analysis | | | | | |
|------|-------|-----|-------|------|-----|-------|------|-----|-------|------|-----|-------|------|
| | | Wald-1 Test | | | Generalized Logistic Regression | | | Wald-1 Test | | | Generalized Logistic Regression | | |
| | | All | NUDIF | UDIF | All | NUDIF | UDIF | All | NUDIF | UDIF | All | NUDIF | UDIF |
| 1 | 3PL | 9.40 | 8.00 | 0.90 | 69.26** | 8.27 | 3.64 | 7.50 | 5.40 | 1.00 | 83.87** | 75.54** | 4.24 |
| 2 | 3PL | 18.65** | 17.45** | 1.00 | 57.01** | 31.36** | 8.43 | 16.97** | 16.00** | 0.80 | 64.24** | 49.11** | 7.24 |
| 3 | 3PL | 19.70 | 18.15 | 0.45 | 183.12** | 59.84** | 8.78 | 13.90 | 12.73 | 0.43 | 216.30** | 194.89** | 11.17 |
| 4 | 2PL | 8.20* | 4.65 | 3.55 | 13.58 | 6.55 | 3.62 | 5.93 | 3.40 | 2.50 | 14.36 | 11.71 | 6.95 |
| 5 | 2PL | 16.15** | 13.95** | 2.20 | 62.00** | 43.37** | 21.57** | 11.87 | 9.60 | 2.30 | 69.23** | 55.22** | 25.64** |
| 6 | 3PL | 5.60 | 3.60 | 1.90 | 18.90* | 3.45 | 1.08 | 5.30 | 3.07 | 1.43 | 25.24** | 22.83** | 0.69 |
| 7 | 3PL | 14.35 | 13.55 | 0.45 | 64.56** | 27.91** | 2.57 | 10.73 | 9.90 | 0.57 | 75.96** | 65.59** | 4.47 |
| 8 | 3PL | 14.70 | 13.80 | 0.05 | 295.18** | 65.62** | 33.86** | 10.33 | 9.33 | 0.43 | 341.97** | 299.76** | 48.36** |
| 9 | 3PL | 1.85 | 1.30 | 0.55 | 5.04 | 3.29 | 5.50 | 1.87 | 1.43 | 0.40 | 6.58 | 2.17 | 5.83 |
| 10 | 3PL | 3.90 | 3.85 | 0.05 | 12.28 | 6.17 | 0.97 | 3.00 | 2.83 | 0.17 | 12.42 | 12.01 | 0.75 |
| 11 | 3PL | 3.35 | 0.75 | 1.65 | 6.79 | 3.83 | 1.42 | 2.50 | 0.70 | 1.13 | 9.62 | 4.56 | 2.87 |
| 12 | 2PL | 0.35 | 0.15 | 0.20 | 1.38 | 9.59 | 3.28 | 1.80 | 1.17 | 0.60 | 12.01 | 2.68 | 9.54 |
| 13 | 3PL | 1.80 | 1.00 | 0.20 | 11.01 | 8.18 | 3.11 | 1.70 | 0.70 | 0.17 | 15.22 | 10.29 | 2.52 |
| 14 | 3PL | 2.20 | 0.60 | 1.00 | 2.62 | 6.25 | 4.89 | 1.70 | 0.47 | 0.67 | 3.32 | 0.92 | 3.84 |
| 15 | 3PL | 5.60 | 4.90 | 0.35 | 18.12* | 21.16* | 3.92 | 4.40 | 3.63 | 0.47 | 15.90 | 12.66 | 5.36 |
| 16 | 3PL | 8.05 | 6.60* | 0.65 | 17.68* | 32.68** | 2.97 | 8.90* | 7.07* | 1.30 | 43.18** | 40.60** | 6.35 |
| 17 | 2PL | 18.30 | 17.80 | 0.50 | 94.98** | 15.79 | 2.53 | 14.87 | 14.53* | 0.33 | 141.45** | 132.77** | 2.09 |
| 18 | 2PL | 5.75 | 3.80 | 2.00 | 59.52** | 17.97 | 6.96 | 6.33 | 3.87 | 2.47 | 67.91** | 59.64** | 10.97 |
| 19 | 2PL | 9.45* | 9.40* | 0.05 | 37.01** | 4.00 | 3.50 | 9.17* | 9.10* | 0.03 | 39.02** | 43.12** | 3.03 |
| 20 | 3PL | . | . | . | 1.14 | 11.44 | 1.67 | . | . | . | 2.19 | 1.40 | 1.85 |
| 21 | 2PL | 1.75 | 0.30 | 1.45 | 0.71 | 22.78* | 1.53 | 1.70 | 0.20 | 1.47 | 2.34 | 0.78 | 0.99 |
| 22 | 2PL | 0.95 | 0.55 | 0.40 | 5.00 | 18.31 | 0.75 | 0.93 | 0.57 | 0.37 | 7.23 | 7.83 | 0.67 |
| 23 | 3PL | 1.90 | 0.10 | 1.30 | 3.07 | 28.81** | 7.07 | 1.63 | 0.10 | 1.20 | 5.38 | 2.05 | 6.53 |
| 24 | 2PL | 11.40 | 8.90 | 2.55 | 36.33** | 16.36 | 5.99 | 8.17 | 6.43 | 1.70 | 75.55** | 63.25** | 7.31 |
| 25 | 3PL | 6.35 | 5.20 | 0.45 | 22.77** | 12.65 | 5.31 | 8.97 | 7.97 | 0.50 | 50.20** | 42.64** | 7.66 |
| 26 | 3PL | 10.95 | 10.10* | 0.60 | 58.63** | 9.89 | 7.95 | 9.87 | 8.37* | 1.23 | 70.83** | 52.36** | 12.49 |
| 27 | 2PL | 2.25 | 0.95 | 1.20 | 13.36 | 8.91 | 5.32 | 1.93 | 1.07 | 0.83 | 14.38 | 11.30 | 5.27 |
| 28 | 2PL | 3.25 | 2.85 | 0.40 | 25.44** | 5.74 | 2.26 | 2.57 | 2.23 | 0.33 | 27.24** | 24.91** | 2.41 |

*Note.* * $p < .05$; ** $p < .01$.

**Table 4**

*TIMSS 2015: DIF Statistics from Two Methods (Six Groups)*

| Item | Model | 6-group analysis | | | | | |
|------|-------|------|-------|------|------|-------|------|
| | | Wald-1 Test | | | Wald-1 Test | | |
| | | All | NUDIF | UDIF | All | NUDIF | UDIF |
| 1 | 3PL | 12.50 | 10.82 | 0.98 | 304.38** | 294.31** | 8.27 |
| 2 | 3PL | 19.68** | 18.14** | 1.42 | 131.36** | 116.98** | 31.36** |
| 3 | 3PL | 11.36 | 8.34 | 1.26 | 237.22** | 206.23** | 59.84** |
| 4 | 2PL | 7.78 | 6.08 | 1.72 | 117.27** | 110.52** | 6.55 |
| 5 | 2PL | 9.28 | 7.34 | 1.92 | 172.62** | 135.12** | 43.37** |
| 6 | 3PL | 3.50 | 1.90 | 1.04 | 41.04** | 33.49** | 3.45 |
| 7 | 3PL | 7.34 | 6.52 | 0.62 | 90.80** | 74.19** | 27.91** |
| 8 | 3PL | 7.04 | 5.76 | 0.30 | 340.64** | 309.47** | 65.62** |
| 9 | 3PL | 1.26 | 0.84 | 0.36 | 4.25 | 1.51 | 3.29 |
| 10 | 3PL | 2.84 | 2.52 | 0.18 | 68.90** | 59.62** | 6.17 |
| 11 | 3PL | 2.60 | 1.42 | 0.68 | 34.75** | 22.86** | 3.83 |
| 12 | 2PL | 3.66 | 2.70 | 0.94 | 80.29** | 63.33** | 9.59 |
| 13 | 3PL | 1.46 | 0.44 | 0.38 | 12.87 | 6.15 | 8.18 |
| 14 | 3PL | 2.16 | 0.66 | 1.10 | 14.35 | 6.07 | 6.25 |
| 15 | 3PL | 4.16 | 3.48 | 0.50 | 43.17** | 25.21** | 21.16* |
| 16 | 3PL | 8.62 | 7.10* | 1.20 | 73.21** | 53.31** | 32.68** |
| 17 | 2PL | 21.12* | 20.22* | 0.90 | 289.68** | 279.83** | 15.79 |
| 18 | 2PL | 6.38 | 5.14* | 1.24 | 123.23** | 107.98** | 17.97 |
| 19 | 2PL | 7.00 | 6.84* | 0.16 | 50.50** | 42.76** | 4.00 |
| 20 | 3PL | . | . | . | 5.45 | 2.89 | 11.44 |
| 21 | 2PL | 1.06 | 0.34 | 0.70 | 28.53* | 17.99* | 22.78* |
| 22 | 2PL | 2.28 | 1.44 | 0.84 | 36.87** | 26.30** | 18.31 |
| 23 | 3PL | 4.16 | 0.82 | 1.78 | 16.29 | 9.60 | 28.81** |
| 24 | 2PL | 23.88 | 22.92 | 0.94 | 354.78** | 346.30** | 16.36 |
| 25 | 3PL | 6.22 | 5.14 | 0.40 | 63.45** | 55.66** | 12.65 |
| 26 | 3PL | 13.40 | 12.36* | 0.86 | 139.79** | 130.61** | 9.89 |
| 27 | 3PL | 2.06 | 1.08 | 1.00 | 23.09 | 17.70* | 8.91 |
| 28 | 2PL | 2.38 | 1.82 | 0.54 | 35.06** | 29.29** | 5.74 |

*Note.* * $p < .05$; ** $p < .01$.

**Table 5**

*IRT-based Model Parameter Estimates and Standard Errors of the Items with DIF (from the Wald-1 Test)*

| Item | 3-group analysis | | | 4-group analysis | | | | 6-group analysis | | | | | |
|---|---|---|---|---|---|---|---|---|---|---|---|---|---|
| | **a1** | **a2** | **a3** | **a1** | **a2** | **a3** | **a4** | **a1** | **a2** | **a3** | **a4** | **a5** | **a6** |
| 2 | 1.04 (0.22) | 1.41 (0.36) | 1.52 (0.35) | 1.04 (0.22) | 1.41 (0.36) | 1.53 (0.35) | 1.34 (0.36) | 1.05 (0.22) | 1.46 (0.38) | 1.60 (0.37) | 1.38 (0.37) | 2.50 (0.83) | 1.77 (0.64) |
| 4 | 1.25 (0.16) | 0.73 (0.16) | 0.92 (0.19) | | | | | | | | | | |
| 5 | 1.43 (0.18) | 1.68 (0.36) | 2.57 (0.54) | | | | | | | | | | |
| 16 | | | | 1.58 (0.29) | 1.64 (0.46) | 2.79 (0.81) | 2.89 (0.89) | | | | | | |
| 17 | | | | | | | | 0.91 (0.15) | 0.82 (0.19) | 1.22 (0.27) | 0.95 (0.25) | 1.51 (0.41) | 1.53 (0.51) |
| 19 | 1.43 (0.18) | 1.37 (0.29) | 1.33 (0.27) | 1.43 (0.18) | 1.36 (0.29) | 1.34 (0.28) | 1.46 (0.35) | | | | | | |
| | **b1** | **b2** | **b3** | **b1** | **b2** | **b3** | **b4** | **b1** | **b2** | **b3** | **b4** | **b5** | **b6** |
| 2 | 0.41 (0.21) | -0.39 (0.22) | -0.27 (0.19) | 0.41 (0.21) | -0.39 (0.22) | -0.27 (0.19) | -0.21 (0.22) | 0.41 (0.21) | -0.36 (0.21) | -0.25 (0.18) | 1.38 (0.37) | -0.75 (0.23) | -1.20 (0.35) |
| 4 | -0.57 (0.10) | -0.27 (0.17) | -0.38 (0.15) | | | | | | | | | | |
| 5 | 0.76 (0.10) | 1.33 (0.31) | 1.08 (0.22) | | | | | | | | | | |
| 16 | | | | 0.08 (0.15) | 0.83 (0.23) | 0.62 (0.17) | 0.68 (0.21) | | | | | | |
| 17 | | | | | | | | 2.08 (0.31) | 0.92 (0.26) | 1.90 (0.40) | 2.74 (0.71) | -0.07 (0.19) | 0.05 (0.24) |
| 19 | 0.04 (0.09) | 0.65 (0.19) | 0.41 (0.16) | 0.04 (0.09) | 0.65 (0.19) | 0.41 (0.15) | 0.54 (0.20) | | | | | | |
| | **c1** | **c2** | **c3** | **c1** | **c2** | **c3** | **c4** | **c1** | **c2** | **c3** | **c4** | **c5** | **c6** |
| 2 | -0.43 (0.26) | 0.55 (0.25) | 0.41 (0.25) | -0.43 (0.26) | 0.55 (0.25) | 0.42 (0.25) | 0.29 (0.27) | -0.43 (0.26) | 0.53 (0.25) | 0.40 (0.25) | 0.27 (0.27) | 1.88 (0.52) | 2.12 (0.46) |
| 4 | . | . | . | | | | | | | | | | |
| 5 | . | . | . | | | | | | | | | | |
| 16 | | | | -0.13 (0.25) | -1.36 (0.42) | -1.72 (0.59) | -1.95 (0.64) | | | | | | |
| 17 | | | | | | | | . | . | . | . | . | . |
| 19 | -0.06 (0.13) | -0.90 (0.19) | -0.55 (0.18) | -0.06 (0.13) | -0.89 (0.19) | -0.55 (0.18) | -0.79 (0.22) | | | | | | |

**Table 6**

*Group-Specific Parameter Estimates and Standard Errors of DIF Items (from Generalized Logistic Regression Procedure)*

| Item | 3-group analysis | | | | 4-group analysis | | | | | | 6-group analysis | | | | | | | | | |
|---|---|---|---|---|---|---|---|---|---|---|---|---|---|---|---|---|---|---|---|---|
| | $\alpha_1$ | $\alpha_2$ | $\beta_1$ | $\beta_2$ | $\alpha_1$ | $\alpha_2$ | $\alpha_3$ | $\beta_1$ | $\beta_2$ | $\beta_3$ | $\alpha_1$ | $\alpha_2$ | $\alpha_3$ | $\alpha_4$ | $\alpha_5$ | $\beta_1$ | $\beta_2$ | $\beta_3$ | $\beta_4$ | $\beta_5$ |
| 1 | 0.07 | 0.64 | -0.02 | 0.05 | -0.02 | 0.59 | -0.53 | 0.00 | 0.06 | 0.09 | 0.29 | 0.95 | -0.15 | 0.69 | 1.96 | -0.11 | 0.00 | 0.05 | 0.30 | 0.01 |
| | (0.45) | (0.43) | (0.07) | (0.07) | (0.42) | (0.40) | (0.47) | (0.06) | (0.06) | (0.07) | (0.46) | (0.43) | (0.50) | (0.47) | (0.43) | (0.11) | (0.11) | (0.12) | (0.15) | (0.13) |
| 2 | 0.26 | 0.01 | 0.10 | 0.17 | 0.32 | 0.10 | 0.27 | 0.08 | 0.15 | 0.07 | 0.16 | 0.10 | 0.53 | -0.10 | 0.83 | 0.13 | 0.24 | 0.03 | 0.43 | 0.28 |
| | (0.40) | (0.41) | (0.07) | (0.08) | (0.37) | (0.38) | (0.39) | (0.06) | (0.06) | (0.07) | (0.39) | (0.39) | (0.39) | (0.42) | (0.39) | (0.11) | (0.11) | (0.11) | (0.14) | (0.14) |
| 3 | 0.30 | 1.19 | 0.02 | 0.09 | 0.35 | 1.16 | -0.59 | 0.02 | 0.10 | 0.14 | 0.20 | 1.47 | -0.21 | 1.83 | 2.27 | 0.03 | 0.08 | 0.13 | -0.38 | -0.32 |
| | (0.55) | (0.51) | (0.09) | (0.08) | (0.52) | (0.49) | (0.59) | (0.07) | (0.07) | (0.08) | (0.57) | (0.51) | (0.61) | (0.52) | (0.50) | (0.13) | (0.12) | (0.14) | (0.14) | (0.13) |
| 4 | | | | | | | | | | | -0.37 | -0.49 | -0.61 | 0.11 | 0.55 | -0.01 | 0.01 | 0.10 | 0.28 | -0.02 |
| | | | | | | | | | | | (0.37) | (0.37) | (0.40) | (0.39) | (0.36) | (0.10) | (0.11) | (0.11) | (0.14) | (0.12) |
| 5 | -4.71 | -1.70 | 0.54 | 0.11 | -4.04 | -1.66 | -2.86 | 0.41 | 0.11 | 0.39 | -6.40 | -2.14 | -2.24 | -2.46 | 0.89 | 1.14 | 0.27 | 0.48 | 0.61 | -0.05 |
| | (1.09) | (0.81) | (0.16) | (0.12) | (0.96) | (0.75) | (0.88) | (0.12) | (0.10) | (0.12) | (1.18) | (0.76) | (0.81) | (0.95) | (0.60) | (0.25) | (0.17) | (0.19) | (0.23) | (0.16) |
| 6 | 0.02 | 0.38 | -0.08 | -0.03 | -0.01 | 0.34 | 0.05 | -0.07 | -0.01 | -0.07 | -0.44 | 0.09 | -0.46 | -1.10 | -0.09 | -0.01 | 0.05 | 0.04 | 0.33 | 0.03 |
| | (0.38) | (0.40) | (0.08) | (0.08) | (0.36) | (0.38) | (0.38) | (0.07) | (0.07) | (0.07) | (0.37) | (0.37) | (0.39) | (0.41) | (0.37) | (0.11) | (0.12) | (0.12) | (0.15) | (0.14) |
| 7 | 0.18 | 0.65 | 0.00 | 0.11 | 0.07 | 0.59 | -0.33 | 0.03 | 0.12 | 0.12 | 0.02 | 0.80 | -0.29 | 0.32 | 0.72 | 0.00 | 0.09 | 0.16 | -0.06 | -0.32 |
| | (0.42) | (0.44) | (0.08) | (0.09) | (0.39) | (0.40) | (0.43) | (0.07) | (0.08) | (0.08) | (0.40) | (0.40) | (0.44) | (0.40) | (0.39) | (0.12) | (0.13) | (0.13) | (0.13) | (0.12) |
| 8 | 0.53 | 0.81 | -0.08 | 0.20 | 0.54 | 0.76 | 0.41 | -0.07 | 0.20 | -0.05 | 0.51 | 0.94 | 0.47 | 1.46 | 1.62 | -0.13 | 0.28 | -0.10 | -0.18 | -0.20 |
| | (0.47) | (0.45) | (0.08) | (0.08) | (0.44) | (0.42) | (0.47) | (0.07) | (0.07) | (0.07) | (0.48) | (0.45) | (0.49) | (0.44) | (0.44) | (0.12) | (0.12) | (0.12) | (0.12) | (0.12) |
| 10 | | | | | | | | | | | -0.98 | -0.37 | -0.77 | 0.41 | 0.88 | 0.10 | -0.03 | 0.14 | 0.06 | -0.25 |
| | | | | | | | | | | | (0.52) | (0.48) | (0.54) | (0.48) | (0.46) | (0.13) | (0.13) | (0.14) | (0.15) | (0.14) |
| 11 | | | | | | | | | | | -0.14 | -0.13 | -0.39 | -0.46 | 0.01 | 0.00 | 0.10 | 0.16 | 0.38 | 0.09 |
| | | | | | | | | | | | (0.35) | (0.35) | (0.38) | (0.39) | (0.36) | (0.10) | (0.10) | (0.11) | (0.13) | (0.12) |
| 12 | | | | | | | | | | | -0.65 | -0.66 | -1.22 | -2.06 | -1.65 | 0.14 | 0.21 | 0.46 | 0.43 | 0.24 |
| | | | | | | | | | | | (0.37) | (0.38) | (0.43) | (0.45) | (0.43) | (0.11) | (0.11) | (0.14) | (0.14) | (0.13) |
| 15 | 0.55 | -0.32 | -0.14 | -0.02 | | | | | | | 0.05 | -0.63 | -0.12 | -0.61 | -0.78 | -0.10 | 0.05 | -0.02 | 0.13 | 0.43 |
| | (0.38) | (0.39) | (0.07) | (0.07) | | | | | | | (0.37) | (0.38) | (0.39) | (0.40) | (0.42) | (0.10) | (0.11) | (0.11) | (0.13) | (0.15) |
| 16 | -0.14 | -0.03 | -0.04 | -0.09 | -0.18 | -0.05 | -1.41 | -0.03 | -0.07 | 0.09 | -0.45 | -0.36 | -1.59 | -0.09 | 0.17 | -0.02 | -0.04 | 0.19 | -0.07 | -0.39 |
| | (0.41) | (0.41) | (0.07) | (0.07) | (0.38) | (0.38) | (0.46) | (0.06) | (0.06) | (0.07) | (0.40) | (0.39) | (0.48) | (0.41) | (0.41) | (0.11) | (0.11) | (0.13) | (0.12) | (0.12) |
| 17 | -0.74 | 1.61 | 0.07 | -0.08 | -0.38 | 1.71 | -0.49 | 0.02 | -0.08 | -0.03 | -1.93 | 1.66 | -0.52 | 1.66 | 1.14 | 0.33 | -0.14 | -0.04 | 0.06 | 0.14 |
| | (0.81) | (0.64) | (0.12) | (0.10) | (0.75) | (0.61) | (0.86) | (0.10) | (0.08) | (0.11) | (0.91) | (0.64) | (0.91) | (0.66) | (0.70) | (0.20) | (0.15) | (0.20) | (0.17) | (0.18) |
| 18 | 1.25 | -0.63 | -0.16 | 0.03 | 1.12 | -0.61 | 0.76 | -0.12 | 0.03 | -0.06 | 1.07 | -0.50 | 0.57 | 0.01 | -0.58 | -0.24 | 0.00 | -0.06 | 0.26 | -0.09 |
| | (0.42) | (0.46) | (0.08) | (0.09) | (0.39) | (0.43) | (0.41) | (0.07) | (0.07) | (0.07) | (0.39) | (0.43) | (0.43) | (0.45) | (0.47) | (0.11) | (0.12) | (0.13) | (0.16) | (0.14) |
| 19 | -0.68 | -1.68 | 0.04 | 0.14 | -0.57 | -1.34 | -1.48 | 0.03 | 0.09 | 0.11 | -1.58 | -1.64 | -1.92 | -1.84 | -0.93 | 0.24 | 0.21 | 0.29 | 0.31 | 0.06 |
| | (0.51) | (0.56) | (0.09) | (0.09) | (0.48) | (0.51) | (0.56) | (0.07) | (0.08) | (0.08) | (0.50) | (0.50) | (0.56) | (0.59) | (0.53) | (0.12) | (0.12) | (0.14) | (0.16) | (0.15) |
| 21 | | | | | | | | | | | -0.42 | -0.46 | -0.50 | -1.48 | -1.33 | 0.07 | 0.15 | 0.14 | 0.49 | 0.27 |
| | | | | | | | | | | | (0.39) | (0.40) | (0.42) | (0.45) | (0.43) | (0.13) | (0.14) | (0.15) | (0.17) | (0.16) |
| 22 | | | | | | | | | | | -1.31 | -0.29 | -0.82 | -2.59 | -2.61 | 0.21 | 0.05 | 0.11 | 0.56 | 0.39 |
| | | | | | | | | | | | (0.62) | (0.55) | (0.63) | (0.90) | (0.96) | (0.15) | (0.14) | (0.15) | (0.22) | (0.23) |
| 24 | 0.14 | 1.39 | 0.00 | -0.13 | 0.26 | 1.39 | -0.82 | -0.01 | -0.11 | 0.08 | 0.04 | 1.25 | -0.36 | 0.70 | 2.15 | -0.02 | -0.16 | 0.00 | 0.19 | 0.12 |
| | (0.47) | (0.43) | (0.08) | (0.07) | (0.44) | (0.41) | (0.51) | (0.07) | (0.06) | (0.08) | (0.47) | (0.43) | (0.51) | (0.47) | (0.44) | (0.12) | (0.11) | (0.13) | (0.15) | (0.15) |
| 25 | 0.51 | 0.55 | -0.19 | -0.13 | 0.32 | 0.34 | -0.77 | -0.14 | -0.08 | 0.00 | -0.09 | 0.37 | -0.93 | 0.15 | 0.02 | -0.15 | -0.16 | 0.04 | -0.04 | -0.19 |
| | (0.37) | (0.38) | (0.07) | (0.07) | (0.34) | (0.35) | (0.39) | (0.06) | (0.06) | (0.07) | (0.36) | (0.35) | (0.40) | (0.37) | (0.37) | (0.10) | (0.10) | (0.11) | (0.12) | (0.12) |
| 26 | 0.48 | 0.56 | 0.10 | -0.02 | 0.40 | 0.41 | -0.39 | 0.11 | 0.01 | 0.16 | 0.12 | 0.44 | -0.52 | 1.07 | 0.91 | 0.20 | 0.00 | 0.30 | 0.10 | 0.20 |
| | (0.48) | (0.48) | (0.08) | (0.08) | (0.44) | (0.44) | (0.49) | (0.07) | (0.07) | (0.08) | (0.47) | (0.45) | (0.52) | (0.45) | (0.46) | (0.12) | (0.11) | (0.13) | (0.13) | (0.14) |
| 28 | 0.36 | -0.47 | -0.03 | 0.00 | 0.18 | -0.52 | -0.16 | 0.01 | 0.02 | 0.05 | -0.15 | -0.70 | -0.46 | -1.28 | -0.10 | 0.05 | 0.07 | 0.16 | 0.31 | 0.12 |
| | (0.46) | (0.49) | (0.08) | (0.09) | (0.43) | (0.45) | (0.46) | (0.07) | (0.07) | (0.08) | (0.42) | (0.43) | (0.46) | (0.51) | (0.45) | (0.11) | (0.12) | (0.13) | (0.15) | (0.14) |

Figures

**Figure 1**

*Item Characteristic Curves for Items Detected with DIF Using Both the Methods*

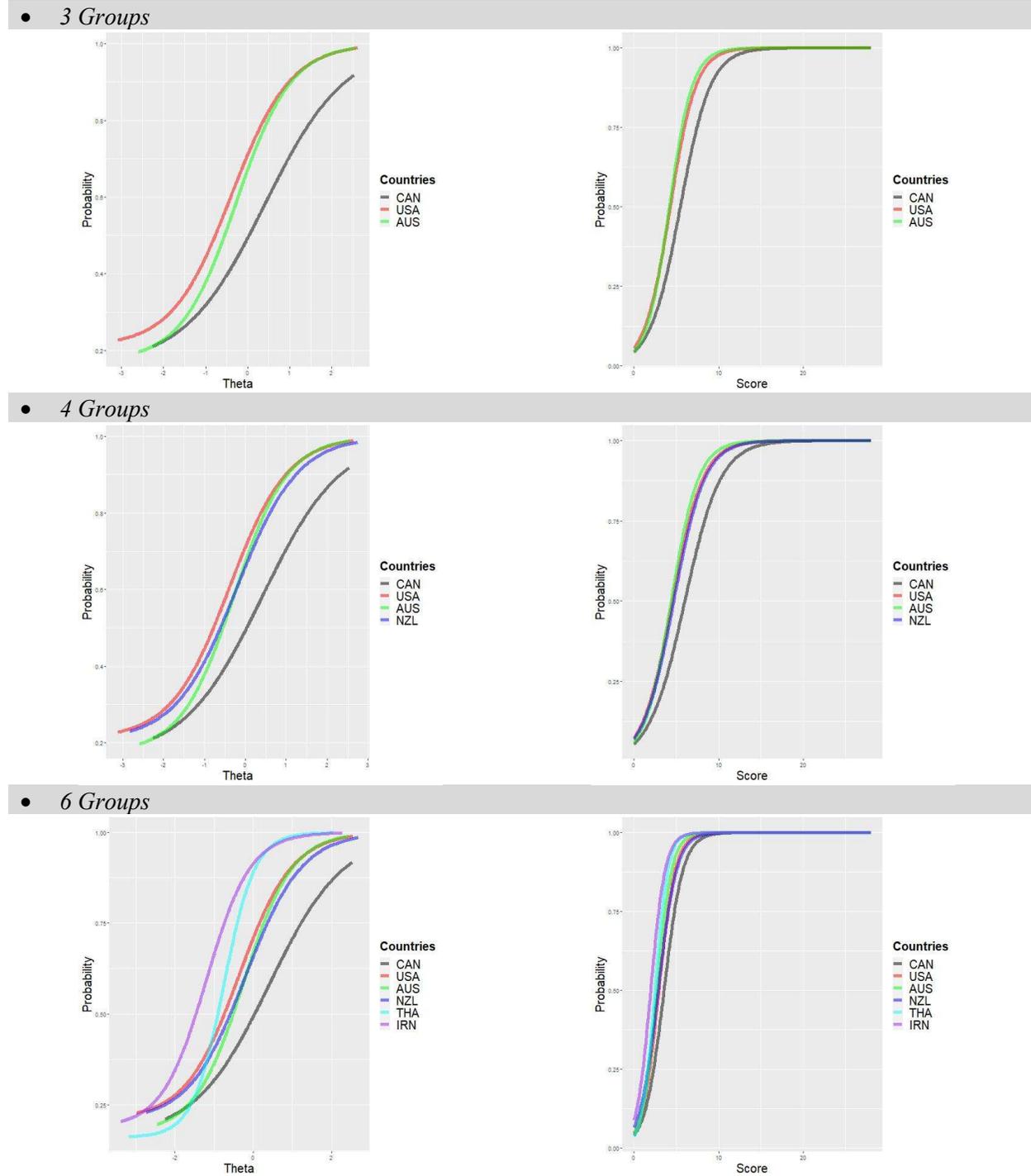

*Note*. Based on data from the Wald-1 test (left) and the generalized logistic regression procedure (right).

**Figure 1 (Continued)**

*Item Characteristic Curves for Items Detected with DIF Using Both the Methods*

<div align="center">

**Item 4**

</div>

- *3 Groups*

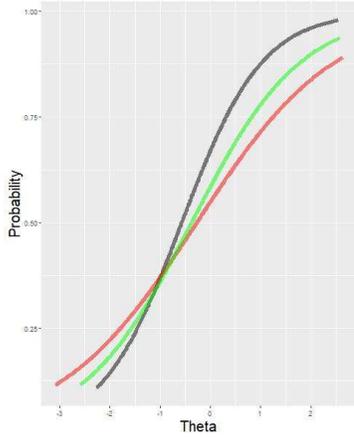 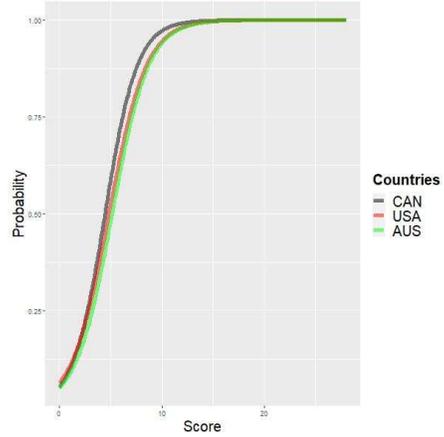

<div align="center">

**Item 5**

</div>

- *3 Groups*

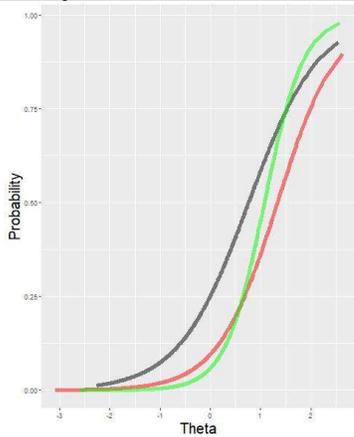 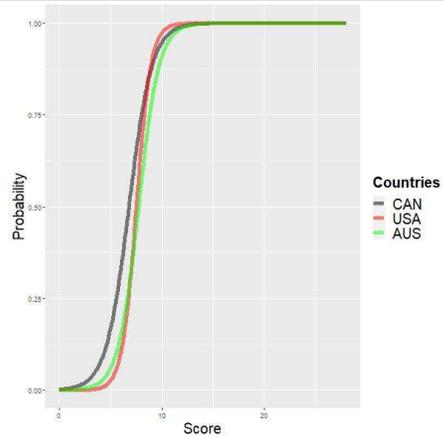

<div align="center">

**Item 16**

</div>

- *4 Groups*

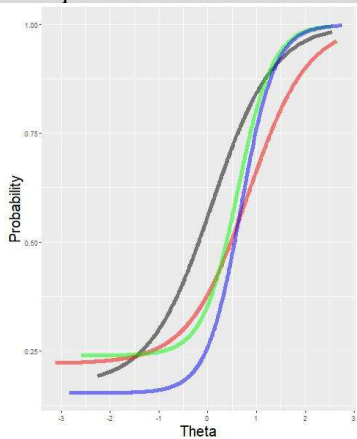 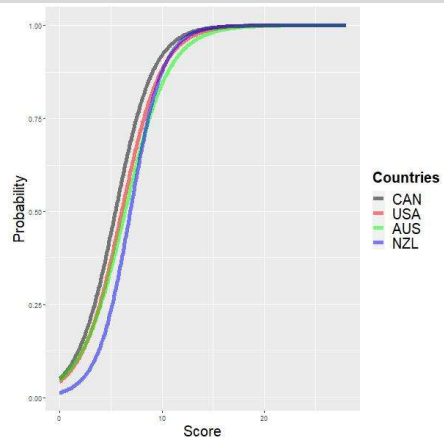

*Note*. Based on data from the Wald-1 test (left) and the generalized logistic regression procedure (right).

**Figure 1 (Continued)**

*Item Characteristic Curves for Items Detected with DIF Using Both the Methods*

### Item 17

- *6 Groups*

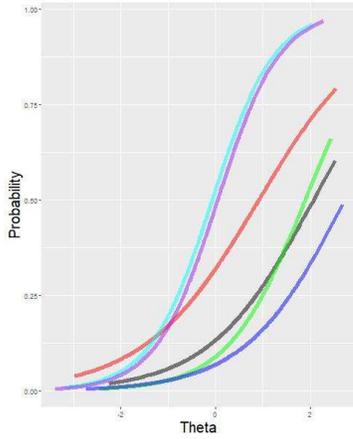 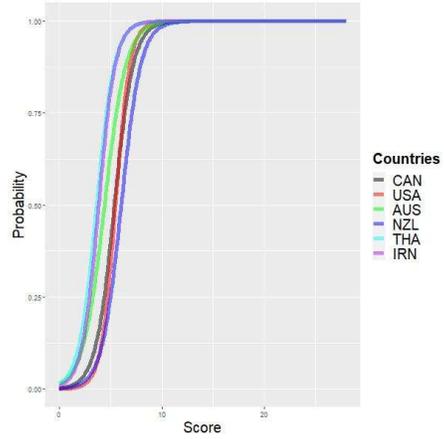

### Item 19

- *3 Groups*

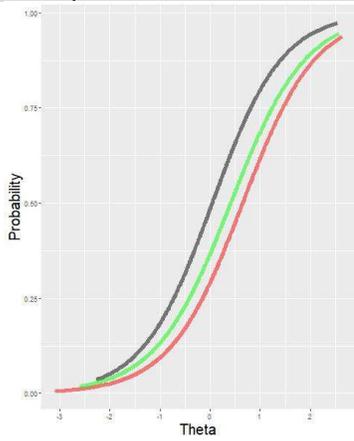 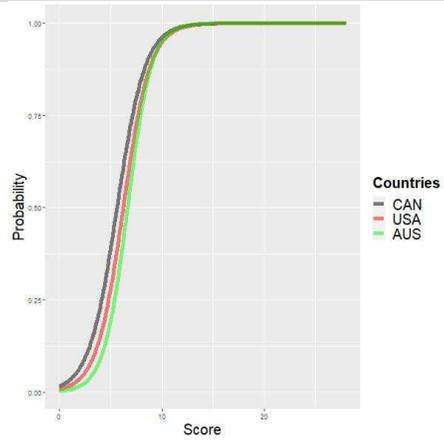

- *4 Groups*

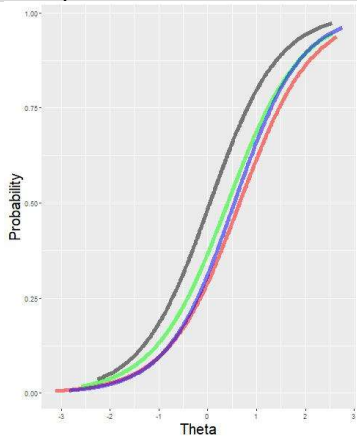 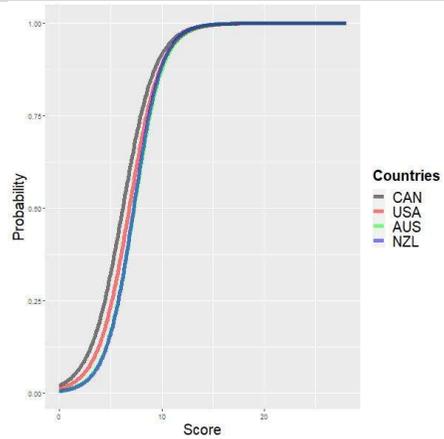

*Note*. Based on data from the Wald-1 test (left) and the generalized logistic regression procedure (right).